\documentclass[aps,prb,twocolumn,groupedaddress]{revtex4-2}

\usepackage{graphicx}
\usepackage{dcolumn}
\usepackage{bm}
\usepackage{verbatim}
\usepackage{ulem}
\usepackage{color}

\begin{document}

\title{On the dilemma between percolation processes and fluctuating pairs as the origin of the enhanced conductivity above the superconducting transition in cuprates}

\author{I. F. Llovo$^{1,2}$}
\author{J. Mosqueira$^{1,2}$}
\email[]{j.mosqueira@usc.es}
\author{F. Vidal$^{1}$}

\affiliation{$^1$ QMatterPhotonics Research Group, Departamento de F\'isica de Part\'iculas, Universidade de Santiago de Compostela, 15782 Santiago de Compostela, Spain}

\affiliation{$^2$ Instituto de Materiais (iMATUS), Universidade de Santiago de Compostela, 15706 Santiago de Compostela, Spain}

\date{\today}

\begin{abstract}
The confrontation between percolation processes and superconducting fluctuations to account for the observed enhanced in-plane electrical conductivity above but near $T_c$ in cuprates is revisited. This dilemma is currently an open and debated question, whose solution would contribute to the phenomenological understanding of the emergence of superconductivity in these compounds. The cuprates studied here, La$_{1.85}$Sr$_{0.15}$CuO$_4$, Bi$_2$Sr$_2$CaCu$_2$O$_{8+\delta}$, and Tl$_2$Ba$_2$Ca$_2$Cu$_3$O$_{10}$, have a different number of superconducting CuO$_2$ ($ab$)-layers per unit-cell length and different Josephson coupling between them, and are optimally-doped to minimize $T_c$-inhomogeneities. The excellent chemical and structural quality of these optimally-doped samples also contribute to minimize the effect of extrinsic $T_c$-inhomogeneities, a crucial aspect when analyzing the possible presence of intrinsic percolative processes. Our analyses also cover the so-called high reduced-temperature region, up to the resistivity rounding onset $\varepsilon_{\rm onset}$. By using the simplest form of the effective-medium theory, we show that possible emergent percolation processes alone cannot account for the measured enhanced conductivity. In contrast, these measurements can be quantitatively explained using the Gaussian-Ginzburg-Landau (GGL) approach for the effect of superconducting fluctuations in layered superconductors, extended to $\varepsilon_{\rm onset}$ by including a total energy cutoff, which takes into account the limits imposed by the Heisenberg uncertainty principle to the shrinkage of the superconducting wavefunction. Our present analysis confirms the adequacy of this cutoff, which was introduced heuristically, and that the effective periodicity length is controlled by the relative Josephson coupling between superconducting layers, two long standing debated aspects of the GGL approaches for multilayered superconductors. These conclusions are reinforced by analyzing, as an example, one of the recent works that allegedly discards the superconducting fluctuations scenario while supporting a percolative scenario for the enhanced conductivity above $T_c$ in cuprates.
\end{abstract}

\maketitle

\section{Introduction}

As early as 1950, Pippard suggested that the electrical resistivity rounding observed around the superconducting transition of low-$T_c$ superconductors could be caused by the interplay between superconducting fluctuations and emergent percolative phenomena associated with chemical and structural inhomogeneities \cite{Pippard50}. Later, as stressed in a review article by Hohenberg \cite{Hohenberg68}, percolative processes were addressed as alternative to the pioneering theoretical works of Ferrell and Schmid \cite{Ferrel67} and, mainly, of Aslamazov and Larkin \cite{Aslamazov68}. These authors explained the resistivity rounding observed above $T_c$, particularly by Shier and Ginsberg \cite{Shier66} and by Glover \cite{Glover67} in amorphous bismuth superconductors, in terms of superconducting fluctuations. The dilemma between superconducting fluctuations and emergent percolation processes was also commented by Kosterlitz and Thouless, when summarizing the studies on the resistivity rounding around $T_c$ in different metallic films \cite{Kosterlitz78}. Some recent examples of this confrontation in low-$T_c$ superconductors and of the influence of disorder and inhomogeneities around $T_c$ on different observables can be seen in Refs.~\onlinecite{Char88,Sacepe08,Venketeswara08,Caprara11,Cohen11,Seibold15,Beutel16,Carbillet16,Sonora-Carballeira19,Ganguly17}.

In cuprate superconductors, the dilemma between percolation processes and superconducting fluctuations was already posed by Bednorz and M\"uller in their seminal work \cite{Bednorz86}. These authors noted that the resistivity decrease observed above but near $T_c$ in their Ba-La-Cu-O samples \textit{results partially from the percolative nature, but possibly also from 2D superconducting fluctuations of double perovskite layers of one of the phases present}. Since then, the relevance of inhomogeneities and percolation processes in the physics of cuprate superconductors was suggested \cite{Phillips90,Phillips03,Kresin06}. This dilemma was addressed at a quantitative level in Ref.~\onlinecite{Maza91} by studying the dc in-plane resistivity, $\rho_{ab}(T)$, of the prototypical optimally-doped YBa$_2$Cu$_3$O$_{7-\delta}$ (YBCO). In that work, the simplest version of the mean-field approach of the effective-medium theory (EMT) \cite{Landauer78,Kirkpatrick73} was used to study possible emergent percolative effects due to $T_c$ inhomogeneities. Additionally, the Lawrence-Doniach (LD) approach for layered superconductors in the Gaussian approximation \cite{Lawrence70,Tsuzuki71,Yamaji72,Tinkham,Larkin05,Skocpol75,Vidal-Ramallo98} was used to analyze the data in terms of fluctuating superconducting pairs created by the unavoidable thermal agitation energy. Ref.~\onlinecite{Maza91} concluded that, in the case of the optimally-doped YBCO, the LD scenario could account for the observed rounding, whereas the possible percolation processes played a negligible role. These conclusions for optimally-doped YBCO have been recently extended to the high reduced-temperature $\varepsilon\equiv\ln(T/T_c)$ region \cite{Llovo-Mosqueira22} by heuristically introducing a total-energy cutoff, which takes into account the quantum localization energy of the short wavelength fluctuating modes \cite{Mosqueira-Carballeira01,Carballeira-Mosqueira01,Rey-Carballeira19,Vidal-Carballeira02,Vidal-Ramallo03}.

Since the earlier results commented above, a number of studies have addressed the influence of possible chemical, structural and electronic disorder on the behavior of different observables around $T_c$ in cuprate superconductors \cite{Mosqueira-Pomar94,Lang94,Casaca97,Shantsev99,Vidal-Veira01,Curras-Ferro03}. However, the dilemma between fluctuating superconducting pairs and percolation processes remains at present an open and debated question, despite the relevance of understanding the emergence of the superconductivity in these materials at a phenomenological level \cite{Llovo-Mosqueira22,Wen02,deMello03,Naqib05,Seto06,Mohapatra06,Benfatto09,Mosqueira-Cabo09,Mosqueira-Dancausa11,Caprara11,Coton-Mercey13,Campi13,Cotón-Ramallo13,Campi15,Naqib15,Zaki17,Seidov18,Pelc18,Popcevic18,Pelc20,Richter21,Borna22}. To contribute to answering that question, detailed $\rho_{ab}(T)$ data, previously measured in high quality crystals and films of La$_{1.85}$Sr$_{0.15}$CuO$_4$, Bi$_2$Sr$_2$CaCu$_2$O$_{8+\delta}$, and Tl$_2$Ba$_2$Ca$_2$Cu$_3$O$_{10}$, will be analyzed here \cite{Curras-Ferro03,Campa92,Maignan94,Vina02}. The studied compounds are also optimally-doped, to minimize the $T_c$-inhomogeneities associated with chemical disorder, but they have different number of $ab$-layers in their periodicity length and different Josephson coupling between them. 

Even in high quality samples the in-plane resistivity rounding close to $T_c$ may be deeply affected by the presence of $T_c$ inhomogeneities, intrinsic or not, mainly in the case of the underdoped samples studied in Ref.~\cite{Popcevic18} and analyzed also in our present paper. So, a considerable and original improvement from previous confrontations between percolation or superconducting fluctuation scenarios is that our analyses also cover the so-called high reduced-temperature region, up to the reduced-temperature onset of the resistivity rounding, $\varepsilon_{\rm onset}\equiv\ln(T_{\rm onset}/T_c)$. In fact, before any detailed comparison with the two scenarios confronted here, the present analyses quantitatively confirm the coincidence for the three studied compounds, well within the experimental uncertainties, of $\varepsilon_{\rm onset}$. In addition, the experimental $\varepsilon_{\rm onset}$ agrees with an estimation based on the limit imposed by the Heisenberg uncertainty principle to the shrinkage of the superconducting wave function well above $T_c$ \cite{Vidal-Carballeira02,Vidal-Ramallo03}. This generalizes the result previously observed in YBa$_2$Cu$_3$O$_{7-\delta}$ \cite{Llovo-Mosqueira22,Rey-Carballeira19} to other optimally-doped cuprates, supporting the fluctuating superconducting pairs scenario.

In this paper, the EMT approach proposed in Ref.~\onlinecite{Maza91} will be used to evaluate the possible presence of emergent percolative processes arising from $T_c$-inhomogeneities to describe the $\rho_{ab}(T)$ rounding above $T_c$. In addition, the adequacy of the fluctuating superconducting pairs scenario will also be tested, for which the LD approach for layered superconductors will be used, extended to the high-$\varepsilon$ region by heuristically including a total-energy cutoff \cite{Mosqueira-Carballeira01,Carballeira-Mosqueira01,Rey-Carballeira19,Vidal-Carballeira02,Vidal-Ramallo03}. Another considerable improvement from previous analyses is that both the effective interlayer distance $d_{\rm eff}$ and the total-energy cutoff were estimated independently. In the case of the highly-anisotropic optimally-doped Bi$_2$Sr$_2$CaCu$_2$O$_{8+\delta}$, and Tl$_2$Ba$_2$Ca$_2$Cu$_3$O$_{10}$, the corresponding analyses could therefore be performed without free parameters. 

Our results confirm that the observed $\rho_{ab}(T)$ rounding around $T_c$ in optimally-doped cuprates can be quantitatively explained by taking into account the unavoidable presence superconducting fluctuations. These analyses also strongly support the adequacy of the total-energy cutoff to extend the LD scenario to high $\varepsilon$, and that $d_{\rm eff}$ is controlled by the relative Josephson coupling between $ab$ layers, two long-standing debated aspects of the GGL approaches \cite{Vidal-Ramallo98,Rey-Carballeira19,Leridon20} ($d_{\rm eff}$ was already studied in the extensions of the LD approach to multilayered superconductors by Maki and Thompson \cite{Maki-Thompson89} and by Klemm \cite{Klemm90}). In addition, the inadequacy of possible percolative effects to explain the enhanced conductivity above $T_c$ in optimally-doped cuprates was also shown. These conclusions enhance the interest of section III.D, where we will briefly examine a recent proposal that questions the conventional GGL scenarios, and that instead suggests percolation processes as the origin of the conductivity enhancement observed just above $T_c$ in cuprates \cite{Popcevic18}.

\section{Methods}
\label{sec:experimental}

\begin{figure*}[t]
\includegraphics[width=0.8\textwidth]{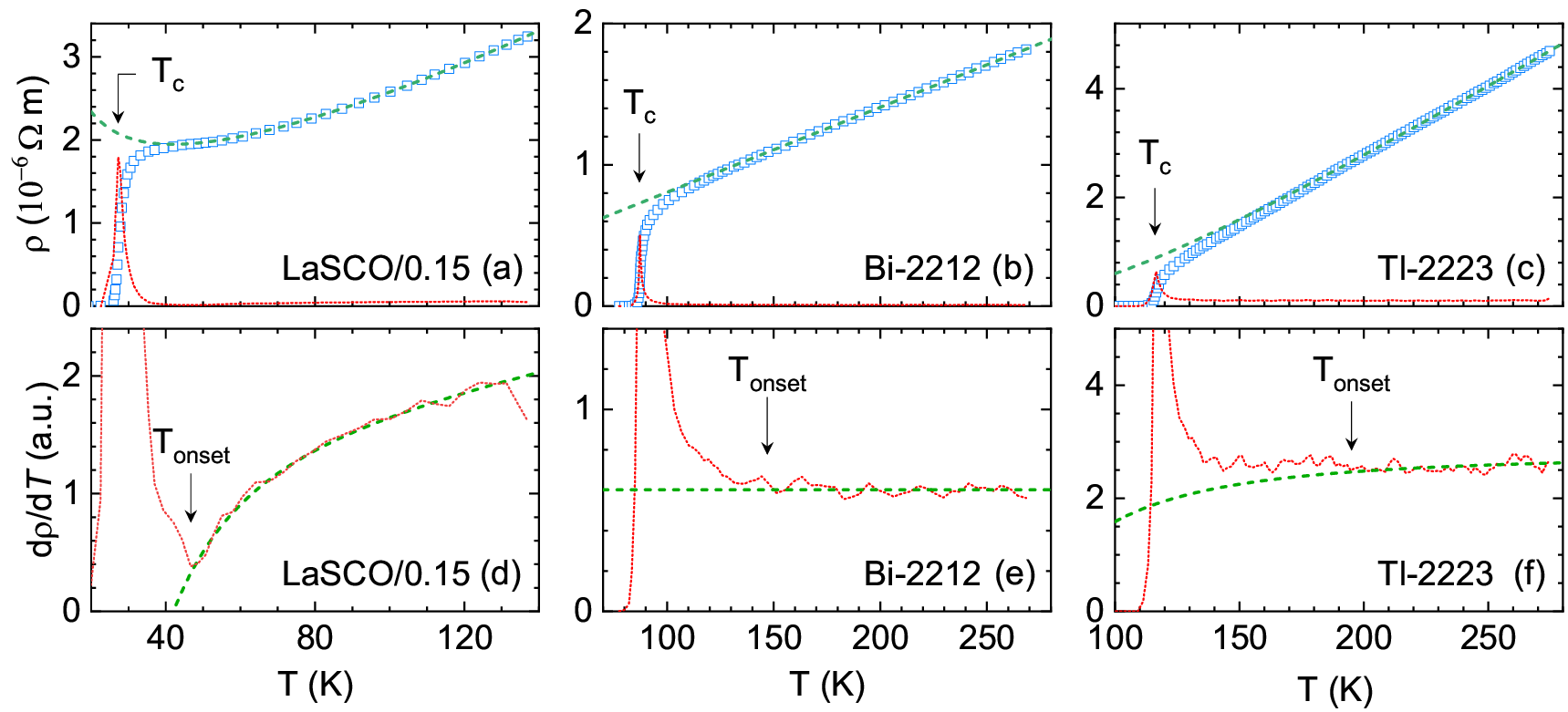}
\caption{Resistive characterization of LaSCO/0.15, Bi-2212 and Tl-2223 (captured from Fig.~2 of Ref.~\onlinecite{Curras-Ferro03} and Fig.~1 of Ref.~\onlinecite{Vina02}). \textbf{a, b, c} In-plane resistivity (blue squares), in-plane background resistivity $\rho_{abB} (T)$ (dashed green line), $d\rho_{ab}(T)/dT$ (dotted red line). \textbf{d, e, f} Detail of $d\rho_{ab}/dT$ and $d\rho_{abB}/dT$ around $T_{\rm onset}$.}  
\label{fig:Fig1}
\end{figure*}

The $\rho_{ab}(T)$ roundings around $T_c$ analyzed here were previously measured in high-quality samples of three different optimally-doped compounds, La$_{1.85}$Sr$_{0.15}$CuO$_4$ (LaSCO/0.15), Bi$_2$Sr$_2$CaCu$_2$O$_{8+\delta}$ (Bi-2212), and Tl$_2$Ba$_2$Ca$_2$Cu$_3$O$_{10}$ (Tl-2223) \cite{Curras-Ferro03,Campa92,Maignan94,Vina02}. These results, presented in Figs.~\ref{fig:Fig1} (a-c), have been chosen due to the exceptional structural and stoichiometric quality of the samples, and to the resolution of the measurements ($\sim1\;\mu\Omega$cm for the resistivity and $\sim10$~mK for the temperature). The uncertainty in the samples geometry and in the distances between electrical contacts leads to an uncertainty in the absolute $\rho_{ab}$ below 20\%. These experimental aspects, as well as the chemical, structural and magnetic characterization of the samples, can be seen in Refs.~\onlinecite{Curras-Ferro03,Campa92,Maignan94,Vina02} and references therein.

%
%
\begin{table}[b]
\begin{ruledtabular}
\begin{tabular}{ccccccc}
Sample & $d$ & $N$ & $T_c$ & $\Delta T_c$ & $T_{\rm onset}$ & $\varepsilon_{\rm onset}$ \\
 &  (nm)&  & (K) & (K) & (K) &  \\
\hline
LaSCO/0.15 & 0.66 & 1 & 27.2$\pm$1.0 & 2.2$\pm$1.0 & 46.6$\pm$2.0 & 0.54$\pm$0.06 \\
Bi-2212 & 1.54 & 2 & 87$\pm$2 & 1.5$\pm$0.4 & 147$\pm$13 & 0.53$\pm$0.09 \\
Tl-2223 & 1.78 & 3 & 116$\pm$2 & 3.9$\pm$1.6 & 195$\pm$4 & 0.52$\pm$0.03 
\end{tabular}
\end{ruledtabular}
\caption{General characteristics of the studied compounds. $d$ is the $ab$ layers periodicity length and $N$ is the number of layers in $d$. $T_c$ and $\Delta T_c$ are the superconducting transition temperature and transition width (FWHM), calculated from $d\rho_{ab}/dT$. $T_{\rm onset}$ is the temperature at which $d\rho_{ab}/dT$ rises above the normal state contribution beyond the noise level, and $\varepsilon_{\rm onset}\equiv\ln(T_{\rm onset}/T_c)$.}
\label{tab:Table1}
\end{table}

%
%

A summary of the main parameters of the samples studied in this work is presented in Table~\ref{tab:Table1}, where $N$ is the number of $ab$-layers per periodicity length, $d$. As shown in Figs.~\ref{fig:Fig1}~(a-c), the superconducting transition temperature $T_c$ was estimated from the maximum of the $d\rho_{ab}/dT$ vs. $T$ curve, whereas the transition width $\Delta T_c$ corresponds to the full width at half maximum. The $\Delta T_c$ values, indicative of the sample homogeneity (see also next section), are within those of the best samples of these compounds \cite{Vidal-Ramallo98,Lang94,Casaca97,Shantsev99,Vidal-Veira01,Curras-Ferro03,Wen02,deMello03,Naqib05,Seto06,Mohapatra06,Benfatto09,Mosqueira-Cabo09,Mosqueira-Dancausa11,Caprara11,Coton-Mercey13,Campi13,Cotón-Ramallo13,Campi15,Naqib15,Zaki17,Seidov18,Pelc18,Popcevic18,Pelc20,Richter21,Borna22,Campa92,Maignan94,Maki-Thompson89}, in particular for LaSCO/0.15, one of the most studied cuprate superconductors \cite{Coton-Mercey13}. In fact, the $c$-axis oriented 150 nm thickness thin film used in this work was grown using a procedure specifically aimed at improving its structural and chemical homogeneity \cite{Curras-Ferro03}.

\section{Results and discussion}

\subsection{Resistivity rounding characterization above $\mathbf{T_c}$}

The temperature dependence of the in-plane resistivity of the three samples studied here is shown in Figs.~\ref{fig:Fig1}~(a-c). To characterize the in plane resistivity rounding observed above but near the superconducting transition, we will use the so-called in-plane paraconductivity, already introduced in the pioneering works on these rounding’s effects in low $T_c$ superconductors and defined as \cite{Pippard50,Hohenberg68,Ferrel67,Aslamazov68,Shier66,Glover67,Tinkham,Larkin05,Skocpol75}
\begin{equation}
\Delta\sigma_{ab}=\frac{1}{\rho_{ab}(\varepsilon)}-\frac{1}{\rho_{abB}(\varepsilon)}.
\label{eq:deltasigma}
\end{equation}
Here $\varepsilon\equiv\ln(T/T_c)$ is the reduced temperature and $\rho_{abB}$ is the background or bare in-plane resistivity, i.e., the normal in-plane resistivity if the critical phenomena associated with the presence of the superconducting transition were absent. The separation between critical and normal contributions is unavoidable when studying the behavior of most of the observables around any phase transition, and it is a consequence that the corresponding theoretical approaches calculate only the first contributions \cite{Wilson83}. A central and general hypothesis in doing such a separation is that the non-critical and the critical behaviors are independent.  Therefore, as usual (see e. g., Refs.~\cite{Hohenberg68,Ferrel67,Aslamazov68,Shier66,Glover67,Skocpol75,Vidal-Ramallo98} and references therein), $\rho_{abB}(T)$ can be determined by fitting an adequate function to the measured $\rho_{ab}(T)$ in a temperature region well above $T_c$, well into the normal state, where the critical phenomena are supposed to be absent (obviously, such a sample-dependent background has no direct physical meaning, an irrelevant aspect when extracting the paraconductivity). The procedure to obtain $\rho_{abB}(T)$ in the studied samples (dashed green lines in Fig.~\ref{fig:Fig1}) is described in detail in Refs.~\cite{Curras-Ferro03,Vina02}. As illustrated in Fig.~\ref{fig:Fig1}, the rounding onset temperature $T_{\rm onset}$ is approximated as the temperature at which $d\rho_{abB}/dT$ deviates from $d\rho_{ab}/dT$ beyond the experimental noise level.

The values for each sample of $T_{onset}$, and of the corresponding $\varepsilon_{onset}$, are indicated in Table~\ref{tab:Table1}. A relevant result in Table~\ref{tab:Table1} which is already worth stressing here is the agreement, well within the experimental uncertainties, of the $\varepsilon_{onset}$ values for the  three optimally doped compounds studied here, which also agree with the one shown by the optimally doped YBa$_2$Cu$_3$O$_{7-\delta}$ compound studied elsewhere \cite{Llovo-Mosqueira22,Rey-Carballeira19}. The implications of this result on the physical origin of the enhanced in-plane conductivity above $T_c$ in optimally doped cuprates will be analyzed in Section III.C. But we must also already note that these $\varepsilon_{onset}$ values are consistent with the predictions of the Gaussian Ginzburg-Landau (GGL) approach under a total-energy cutoff \cite{Vidal-Carballeira02,Vidal-Ramallo03}, favoring the presence of fluctuating superconducting pairs as the origin of the precursor conductivity in these superconductors.

%
%
\begin{figure}[h]
\includegraphics[width=0.4\textwidth]{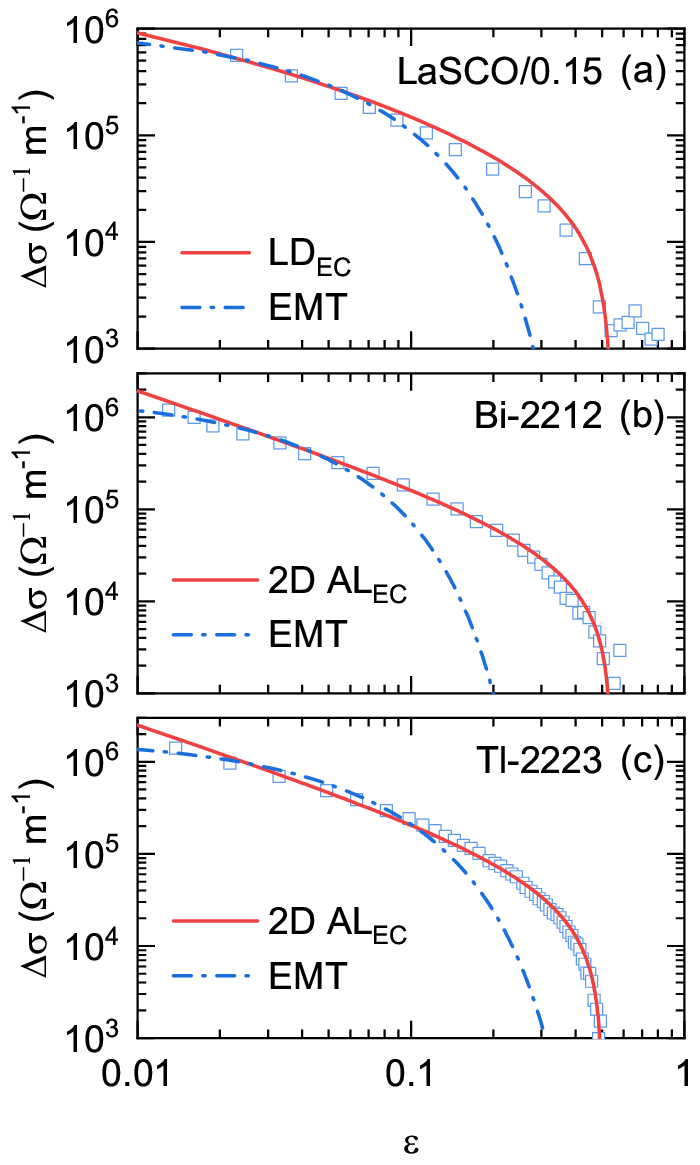}
\caption{Analysis of $\Delta\sigma_{ab}$ (blue squares) of the samples studied. The dot-dashed blue lines were obtained by using the EMT approach (the corresponding parameters are summarized in Table \ref{tab:Table2}). The solid red curves correspond to the fluctuating superconducting pairs scenario and were obtained by using the GGL approach with a total-energy cutoff (the corresponding parameters are summarized in Table \ref{tab:Table3}). The excellent agreement of the GGL curves, obtained without any free parameters for Bi-2212 and Tl-2223, contrasts with the poor results of the percolative model, even at a qualitative level. See the main text for details.}
\label{fig:Fig2}
\end{figure}

The data points in Fig.~\ref{fig:Fig2} are $\Delta\sigma_{ab}(\varepsilon)$ for the three compounds studied here, calculated from Eq.~(\ref{eq:deltasigma}). In the case of Bi-2212 and Tl-2223, for $0.03 \stackrel{<}{_\sim} \varepsilon\stackrel{<}{_\sim} 0.2$, $\Delta\sigma_{ab}\propto\varepsilon^{-1}$, an $\varepsilon$-dependence consistent with 2D superconducting fluctuations as we will see in section III.C. For $\varepsilon\stackrel{<}{_\sim} 0.02$ the data may be affected by the so-called full-critical effects \cite{Mosqueira-Pomar94,Casaca97,Curras-Ferro03,deMello03}. Also, $T_c$-inhomogeneities could affect the data down to $\varepsilon\approx\Delta T_c/T_c$, that is 0.017 and 0.034 for Bi-2212 and Tl-2223, respectively. In the case of LaSCO/0.15, $\Delta\sigma_{ab}(\varepsilon)$ is affected close to $T_c$ by its much larger $\Delta T_c/T_c\approx 0.09$ (which also complicates its $T_c$ estimation), and also presents a somewhat smoother $\varepsilon$-dependence at higher $\varepsilon$. As stressed before, these behaviors close to $T_c$ further enhance the relevance of extending the analyses to the high reduced temperature region, up to $\varepsilon_{\rm onset}$, as it is done in the next two Subsections.

\subsection{Analysis in terms of percolative processes}

To probe if emergent percolative processes associated with the presence of chemical and structural inhomogeneities could be the origin of the paraconductivity, Bruggeman’s effective-medium theory (EMT) \cite{Landauer78,Kirkpatrick73} can be used, as proposed in Ref.~\onlinecite{Maza91}. The main hypothesis is to assume that doping-level inhomogeneities at long length scales (much larger than the superconducting coherence length amplitudes), cause a spatial $T_c$-distribution. The effective in-plane electrical conductivity $\langle\sigma_{ab}\rangle$ can be obtained by numerically solving the implicit equation
\begin{equation}
\int_0^\infty\frac{\sigma_{ab}(T,T_c)-\langle\sigma_{ab}(T)\rangle}{\sigma_{ab}(T,T_c)+2\langle\sigma_{ab}(T)\rangle}G(T_c)dT_c=0,
\label{eq:emt-integral}
\end{equation}
where $G(T_c)$ is the volume fraction of domains with $T_c$ as critical temperature. $G(T_c)$ can be approximated by a Gaussian distribution, characterized by an average critical temperature $T_c^{EMT}$ and a full width at half-maximum (FWHM) $\Delta T_c^{EMT}$. The in-plane electrical conductivity of the different domains is assumed to be

\begin{equation}
\sigma_{ab}(T,T_c)\to\left\{
\begin{array}{ll}\infty & {\rm if}\;T_c^{\rm EMT}>T\\
1/\rho_{abB}(T)& {\rm if}\;T_c^{\rm EMT}<T
\end{array}\right.
\label{eq:emt-sigma}
\end{equation}

\begin{figure*}[t]
\includegraphics[width=0.7\textwidth]{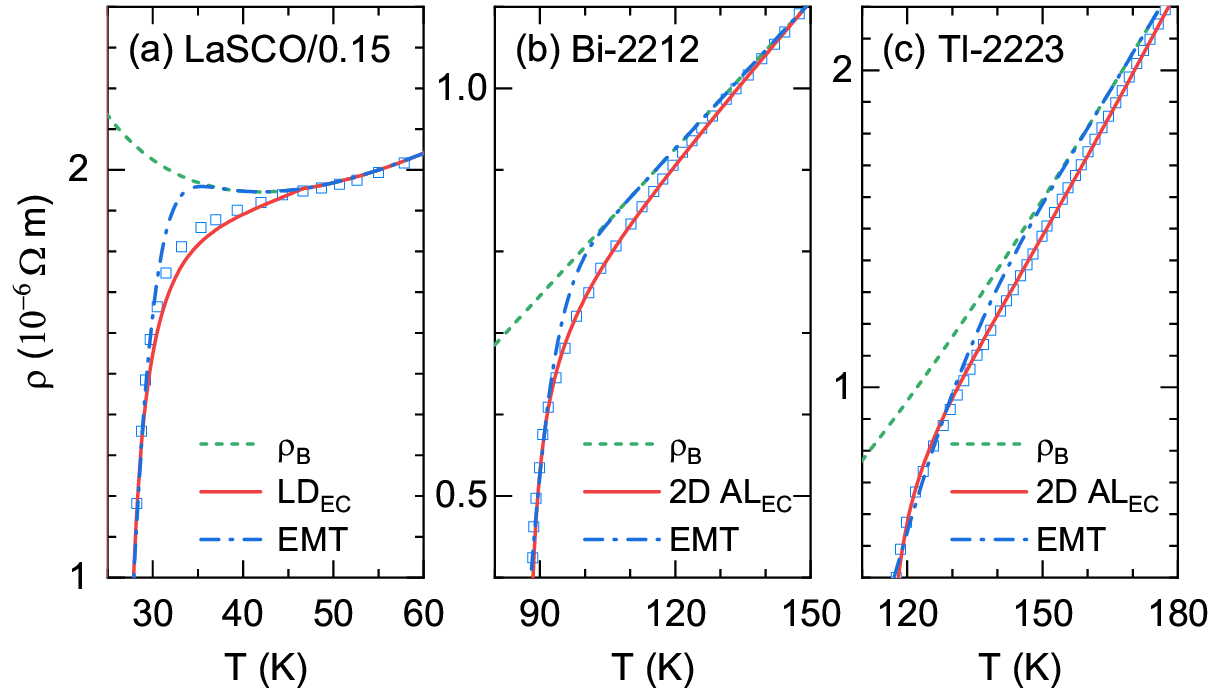}
\caption{Comparison of $\rho_{ab}(T)$ with the EMT approach (dot-dashed blue lines), and the GGL approaches (solid red lines) in the extended mean field region above $T_c$ for \textbf{a} LaSCO/0.15, \textbf{b} Bi-2212 and \textbf{c} Tl-2223. The background resistivity $\rho_B$ is also shown for comparison.}
\label{fig:Fig3}
\end{figure*}

The best fits of Eq.~(\ref{eq:emt-integral}) to the experimental $\Delta\sigma_{ab}(\varepsilon)$ with $\Delta T_c^{\rm EMT}$ and $T_c^{\rm EMT}$ as free parameters are shown in Fig.~\ref{fig:Fig2}. The resulting values for these two parameters are presented in Table \ref{tab:Table2}. The disagreement is dramatic for $\varepsilon > 0.1$, and can also be seen in the $\rho_{ab}(T)$ representation in Fig.~\ref{fig:Fig3}. Notice also that the seeming agreement observed closer to $T_c$ in Fig.~\ref{fig:Fig3} is questioned by the large $\Delta T_c^{\rm EMT}$ values resulting from the EMT fits, a factor $\sim10$ larger than the resistive transition widths for Bi-2212 and Tl-2223 (see Table \ref{tab:Table1}). The latter are in turn consistent with the diamagnetic transition widths observed in similar samples \cite{Vidal-Ramallo98,Rey-Carballeira19,Mosqueira-Dancausa11}, confirming that they are a reasonable indicative of their actual $T_c$-inhomogeneities. In fact, previous paraconductivity analyses on the grounds of percolative scenarios also lead to unphysical $T_c$ distributions, with $\Delta T_c^{\rm EMT}$ values comparable to the $T_c$ of the samples \cite{Maza91,Llovo-Mosqueira22,Pelc18,Popcevic18,Pelc20}.

%
%

%
%
\begin{table}[h]
\begin{ruledtabular}
\begin{tabular}{ccc}
Sample & $T_c^{EMT}$ & $\Delta T_c^{EMT}$  \\ 
& (K) & (K) \\
\hline
 LaSCO/0.15 & 24.5$\pm$0.3 & 8.5$\pm$0.6 \\ 
 Bi-2212 & 80.1$\pm$0.7 & 18.0$\pm$1.5 \\ 
 Tl-2223 & 101.3$\pm$1.0 & 44.2$\pm$2.4 \\ 
\end{tabular}
\end{ruledtabular}
\caption{Parameters obtained in the percolative scenario by fitting Eq.~(\ref{eq:emt-integral}) to the resistivity rounding just above $T_c$ in the studied compounds. The $\Delta T_c^{EMT}$ values are particularly high (comparable to $T_c^{EMT}$) and the fit quality is bad, as it can be seen in Figs. \ref{fig:Fig2} and \ref{fig:Fig3}.}
\label{tab:Table2}
\end{table}

\subsection{Analysis in terms of superconducting fluctuations}

We will now probe if the presence of fluctuating superconducting pairs, created by the thermal agitation energy, could be the primary cause of the resistivity rounding effects observed just above $T_c$ in the three compounds studied here. For that, we take advantage of the well-established absence of appreciable indirect (Maki-Thompson) fluctuation effects on the in-plane paraconductivity in cuprate superconductors \cite{Vidal-Ramallo98,Rey-Carballeira19}. Therefore, the Lawrence-Doniach (LD) approach of the phenomenological Gaussian-Ginzburg-Landau (GGL) scenario can be used \cite{Lawrence70,Tsuzuki71,Yamaji72,Tinkham,Larkin05,Skocpol75}. This approach, well adapted to the layered nature of the cuprate superconductors, can be extended to the high-$\varepsilon$ region by introducing a total-energy cutoff (see below), which leads to \cite{Mosqueira-Carballeira01,Carballeira-Mosqueira01,Rey-Carballeira19,Vidal-Carballeira02,Vidal-Ramallo03}
\begin{equation}
\label{eq:delta_sigma_bld}
\Delta\sigma_{ab}(\varepsilon)=\frac{e^2}{16\hbar d_{\rm eff}}\left[\frac{1}{\varepsilon}\left(1+\frac{B_{LD}}{\varepsilon}\right)^{-\frac{1}{2}}-\frac{1}{\varepsilon^c}\left(1+\frac{B_{LD}}{\varepsilon^c}\right)^{-\frac{1}{2}}\right].
\end{equation}
Here, $B_{LD}\equiv[2\xi_c(0)/d_{\rm eff}]^2$ is the LD coupling parameter, $\xi_c(0)$ the $c$-axis coherence length amplitude, $\varepsilon^c$ the total energy cutoff constant, and $d_{\rm eff}$ the effective distance between $ab$-layers, which depends on their relative Josephson coupling (see below). In the highly anisotropic compounds studied here, the Josephson coupling between neighboring layers can be neglected, i.e., $\xi_c(0)\sim0$ (or $B_{LD}\sim 0$), and Eq.~(\ref{eq:delta_sigma_bld}) reduces to
\begin{equation}
\label{eq:delta_sigma_2D}
\Delta\sigma_{ab}(\varepsilon)=\frac{e^2}{16\hbar d_{\rm eff}}\left(\frac{1}{\varepsilon}-\frac{1}{\varepsilon^c}\right).
\end{equation}
In the absence of cutoff, i.~e., $\varepsilon^c\to\infty$, Eq.~(\ref{eq:delta_sigma_bld}) further reduces to the well-known Aslamazov-Larkin (AL) result for the superconducting fluctuations of layered superconductors (which is the same as the AL result in the 2D limit but with $d_{\rm eff}$ replacing the film thickness, as it is easy to check) \cite{Aslamazov68}. The conventional GGL approach (and so Eqs.~(\ref{eq:delta_sigma_bld}) and (\ref{eq:delta_sigma_2D})) is applicable in the mean field region, which is limited by $\varepsilon_{LG}$, the Levanyuk-Ginzburg reduced-temperature \cite{Tinkham,Larkin05,Skocpol75,Vidal-Ramallo98,Rey-Carballeira19,Ramallo-Pomar96}. Below $\varepsilon_{LG}$, the superconducting fluctuations enter in the so-called full-critical (non-Gaussian) region, and theoretical approaches such as the 3D-XY model must be applied \cite{Ramallo-Pomar96}. Although $\varepsilon_{LG}$ depends on the particular characteristics of each compound, it is of the order of $10^{-2}$ for those studied here \cite{Vidal-Ramallo98,Coton-Mercey13,Ramallo-Pomar96}. For Bi-2212 and Tl-2223, this limit is near the $\Delta T_c/T_c$ values shown in Table~\ref{tab:Table1}, and it is well below the corresponding $\Delta T_c/T_c$ in the case of the LSCO sample. Therefore, to avoid the possible influence of both full-critical fluctuations and $T_c$-inhomogeneities, the comparison with Eqs.~(\ref{eq:delta_sigma_bld}) and (\ref{eq:delta_sigma_2D}) will be restricted to $\varepsilon\stackrel{>}{_\sim}0.02$.

In the high-$\varepsilon$ region, for $\varepsilon\stackrel{>}{_\sim}0.1$, it is also well known that the GGL approach overestimates the contribution of short-wavelength fluctuation modes \cite{Tinkham,Larkin05,Skocpol75,Vidal-Ramallo98}. Trying to overcome this shortcoming, the so-called kinetic-energy or momentum cutoff was early heuristically proposed when analyzing the superconducting fluctuations in low-$T_c$ superconductors \cite{Tinkham,Skocpol75,Johnson78}, and later in cuprate superconductors \cite{Tinkham,Larkin05,Skocpol75,Vidal-Ramallo98,Freitas87,Lee-Klemm89,Gauzzi95}. More recently, this momentum cutoff was recovered by using a microscopic approach based on diagrammatic techniques \cite{Larkin05}. However, such a cutoff does not take into account that, when the temperature increases above $T_c$, the Heisenberg uncertainty principle limits the shrinkage of the superconducting coherence length $\xi(T)$ below $\xi_0$, the Pippard or BCS characteristic length, representative of the Cooper pairs’ size \cite{Vidal-Carballeira02,Vidal-Ramallo03}. To solve this problem, the so-called total-energy cutoff was also heuristically introduced, to include the kinetic energy of the fluctuating modes as well as their quantum localization energy \cite{Mosqueira-Carballeira01,Carballeira-Mosqueira01,Vidal-Carballeira02}. Up to now, that total-energy cutoff has not been recovered by using microscopic approaches (in fact, the difficulties to extend the microscopic calculations to $\Delta\sigma$ at high-$\varepsilon$ were earlier commented, for instance, in Ref.~\onlinecite{Patton72}). However, the heuristic introduction of the total energy cutoff extends the applicability of Eqs.~(\ref{eq:delta_sigma_bld}) and (\ref{eq:delta_sigma_2D}) up to the temperature onset of the fluctuations \cite{Mosqueira-Carballeira01,Carballeira-Mosqueira01,Rey-Carballeira19,Vidal-Carballeira02,Vidal-Ramallo03}, and for $\varepsilon>\varepsilon^c$ (the total-energy cutoff constant) all the fluctuating modes are correctly suppressed. 

As first proposed in Refs.~\onlinecite{Vidal-Carballeira02,Vidal-Ramallo03}, $\varepsilon^c$ can be estimated through the condition $\xi(\varepsilon_{\rm onset})=\xi_0$. As already stressed therein, and later when analyzing the superconducting fluctuations in optimally-doped YBCO \cite{Rey-Carballeira19}, this condition is general and must be applied to any theoretical description in terms of fluctuating superconducting pairs. Assuming the mean-field temperature dependence of the coherence length, $\xi(T)=\xi(0)\varepsilon^{-1/2}$, and the relationship between $\xi_0$ and $\xi(0)$ proposed in the BCS theory, which in the clean limit is \cite{deGennes} $\xi(0)=0.74\xi_0$, the above condition leads to $T_{\rm onset}\approx1.7T_c$, hence $\varepsilon^c\approx0.55$ (in anisotropic superconductors the above expressions are valid for both $ab$ and $c$ directions). The agreement well within the experimental uncertainty of this $\varepsilon^c$ value and the $\varepsilon_{\rm onset}$ observed in the three optimally-doped cuprates studied here is a remarkable result. Such an agreement was also found when analyzing the $\rho_{ab}(T)$ rounding and the precursor diamagnetism above $T_c$ in high-quality YBCO samples \cite{Rey-Carballeira19}, and already suggests that fluctuating superconducting pairs are responsible for the precursor electrical conductivity in optimally-doped cuprates.

Considering the above comments, Eqs.~(\ref{eq:delta_sigma_bld}) and (\ref{eq:delta_sigma_2D}) were used to analyze the paraconductivity data in the $\varepsilon$-region between 0.02 and $\varepsilon_{\rm onset}$. In doing so, an appreciable improvement relative to previous works is the independent estimation of $\varepsilon^c$ and $d_{\rm eff}$. The origin of $\varepsilon^c$ and the $d_{\rm eff}$ estimation are long standing still open issues of the GGL scenario \cite{Rey-Carballeira19,Vidal-Carballeira02,Vidal-Ramallo03,Mosqueira-Pomar94,Lang94,Casaca97,Shantsev99,Vidal-Veira01,Curras-Ferro03,Wen02,deMello03,Naqib05,Seto06,Mohapatra06,Benfatto09,Mosqueira-Cabo09,Mosqueira-Dancausa11,Caprara11,Coton-Mercey13,Leridon20}. For $\varepsilon^c$, we will use $\varepsilon_{\rm onset} \equiv\ln(T_{\rm onset}/T_c)$. Regarding $d_{\rm eff}$, it is now well established that it is controlled by the relative, not the absolute, Josephson coupling between superconducting layers \cite{Vidal-Ramallo98,Rey-Carballeira19,Maki-Thompson89,Klemm90,Ramallo-Pomar96}. Moreover, it was also shown that even for interlayer coupling differences as big as 100, the effective periodicity length can be approximated as $d/N$, where $N$ is the number of layers per periodicity length~$d~$ \cite{Vidal-Ramallo98,Ramallo-Pomar96}. This conclusion can be then applied to the compounds studied here and we will approximate $d_{\rm eff}=d/N$, as summarized in Table \ref{tab:Table2}. Notice that the independent estimations of $\varepsilon^c$ and $d_{\rm eff}$ allows the use of Eq.~(\ref{eq:delta_sigma_bld}) without free parameters to analyze the measurements in Bi-2212 and Tl-2223. 

The solid red line in Fig.~\ref{fig:Fig3}(a) is the best fit of Eq.~(\ref{eq:delta_sigma_bld}) to the~$\Delta\sigma_{ab}(\varepsilon)$ measured in LaSCO/0.15. As previously discussed, this fit was performed in the $\varepsilon$-region between 0.02 and $\varepsilon_{\rm onset} = 0.54$. Moreover, in this single layered compound (with $N=1$) $d_{\rm eff}$ coincides with the $ab$-layers periodicity length, i.e., $d_{\rm eff} = 0.66$ nm, as it is shown in Table~\ref{tab:Table3}. $B_{LD}$ is then the only free parameter, and the best fit leads to $B_{LD}= 0.048\pm0.004$, hence $\xi_c(0)=0.072\pm0.006$ nm. This $\xi_c(0)$ value is much smaller than $d_{\rm eff}$, which is indicative of weak Josephson coupling. Although weak, this interlayer coupling justifies that the critical exponent of the measured in-plane paraconductivity is between -1/2 and -1 for $0.02 \stackrel{<}{_\sim} 0.1$, confirming the 2D-3D behavior of the superconducting fluctuations in this compound \cite{Vidal-Ramallo98,Curras-Ferro03}.

%
%
\begin{table}[t]
\begin{ruledtabular}
\begin{tabular}{cccccc}
Sample  & $d_{\rm eff}$  & $\varepsilon^c$ & $B_{LD}$  & $\xi_c(0)$ & $\xi_{ab}(0)$ \\ 
  & (nm)  &  &   &  (nm) & (nm) \\
\hline
LaSCO/0.15 & 0.66 & 0.54 & 0.048$\pm$0.004 & 0.072$\pm$0.006 & 3.2 \\
Bi-2212 & 0.77 & 0.53 & 0 & $\sim$~0 & 0.9 \\
Tl-2223 & 0.593 & 0.52 & 0 & $\sim$~0 & 1.0 \\ 
\end{tabular}
\end{ruledtabular}
\caption{Parameters of the fluctuating superconducting pair’s scenario, obtained from the GGL analysis of the resistivity roundings just above $T_c$ in the studied optimally-doped compounds. $d_{\rm eff}$ is the effective interlayer distance and $B_{LD}$ the Lawrence-Doniach parameter. The total-energy cutoff constant $\varepsilon_c$ in Eqs.~(\ref{eq:delta_sigma_bld}) and (\ref{eq:delta_sigma_2D}) was set to the value of $\varepsilon_{\rm onset}$ (see Table \ref{tab:Table1}). These different parameters were obtained as indicated in the main text. For completeness, values of the in-plane coherence length amplitude taken from Refs.~\onlinecite{Suzuki91, Mosqueira-Miramontes96,Mosqueira-Cabo07} are also included.}
\label{tab:Table3}
\end{table}

For the multi-layered compounds Bi-2212 ($N=2$, $d=1.54$ nm) and Tl-2223 ($N=3$, $d=1.78$ nm), the effective interlayer distances are $d_{\rm eff} = d/2 = 0.77$ nm and $d_{\rm eff} = d/3 =0.593$ nm, respectively. As shown in Figs.~\ref{fig:Fig1}(e-f), the aforementioned procedure to estimate $\varepsilon_{\rm onset}$ was also followed, obtaining $\varepsilon_{\rm onset} = 0.53$ for Bi-2212, and $\varepsilon_{\rm onset} = 0.52$ for Tl-2223. Moreover, the experimental results in Figs.~\ref{fig:Fig3}(b-c) shows that for $0.01 \lesssim\varepsilon\lesssim0.1$, $\Delta\sigma_{ab}(\varepsilon)\propto \varepsilon^{-1}$ well within the experimental uncertainties, corresponding to the 2D limit consistent with a negligibly small $\xi_c(0)$ value \cite{Vidal-Veira88}. Therefore, when analyzing the corresponding $\Delta\sigma_{ab}(\varepsilon)$ in terms of the GGL approach, Eq.~(\ref{eq:delta_sigma_2D}) was used. The resulting paraconductivity curves, obtained without free parameters, are shown as solid red lines in Figs.~\ref{fig:Fig3}(b-c). As it can be seen, the experimental data are remarkably well explained by the extended GGL approach for multilayered superconductors in the $\varepsilon$-range analyzed.

%
%
\begin{figure}[t]
\includegraphics[width=0.4\textwidth]{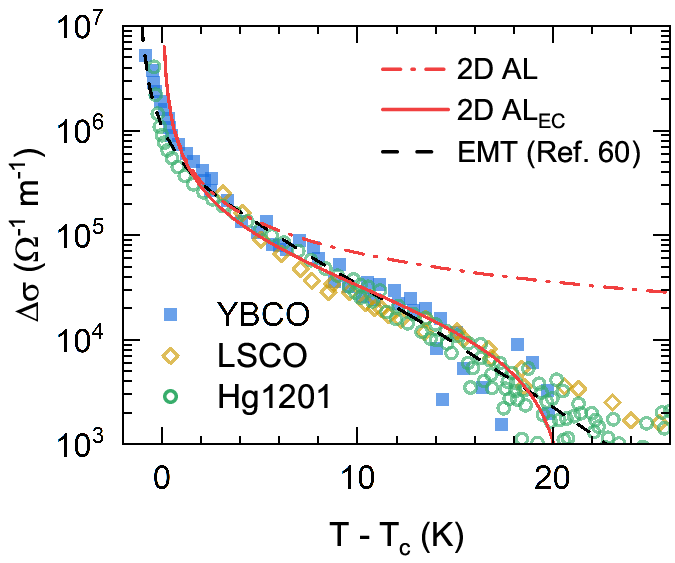}
\caption{Paraconductivity data for YBCO, LSCO and Hg1201, captured from Fig.~2 of Ref.~\onlinecite{Popcevic18}. The data are compared with the percolative model of Ref.~\onlinecite{Popcevic18} (denoted as EMT, dashed black line), the 2D AL result without cutoff (dot dashed red line), and with a total-energy cutoff (2D AL$_{\rm EC}$, solid red line). All these curves correspond to Hg1201 (see main text for details). Data for $\Delta\sigma_{ab}<10^3\;\Omega^{-1}$m$^{-1}$ were not considered due to the high dispersion.}
\label{fig:Fig4}
\end{figure}

\subsection{Comparison with some recent results}
\label{sec:app}

To complement our analyses, here we will analyze one of the recent works that propose a universal percolative scenario for the paraconductivity in cuprate superconductors \cite{Zaki17,Seidov18,Pelc18,Popcevic18,Pelc20,Richter21}, in opposition to the GGL approaches. This example corresponds to Ref.~\onlinecite{Popcevic18}, and we will show that the arguments presented therein against the GGL approach and the ones supporting the percolative scenario are unjustified. The interest of this section is enhanced by the fact that our results can also be applied to the analyses of other authors questioning the GGL scenarios for the rounding effects observed above $T_c$ in cuprates \cite{Pelc18,Pelc20}.

\begin{figure*}[t]
\includegraphics[width=0.75\textwidth]{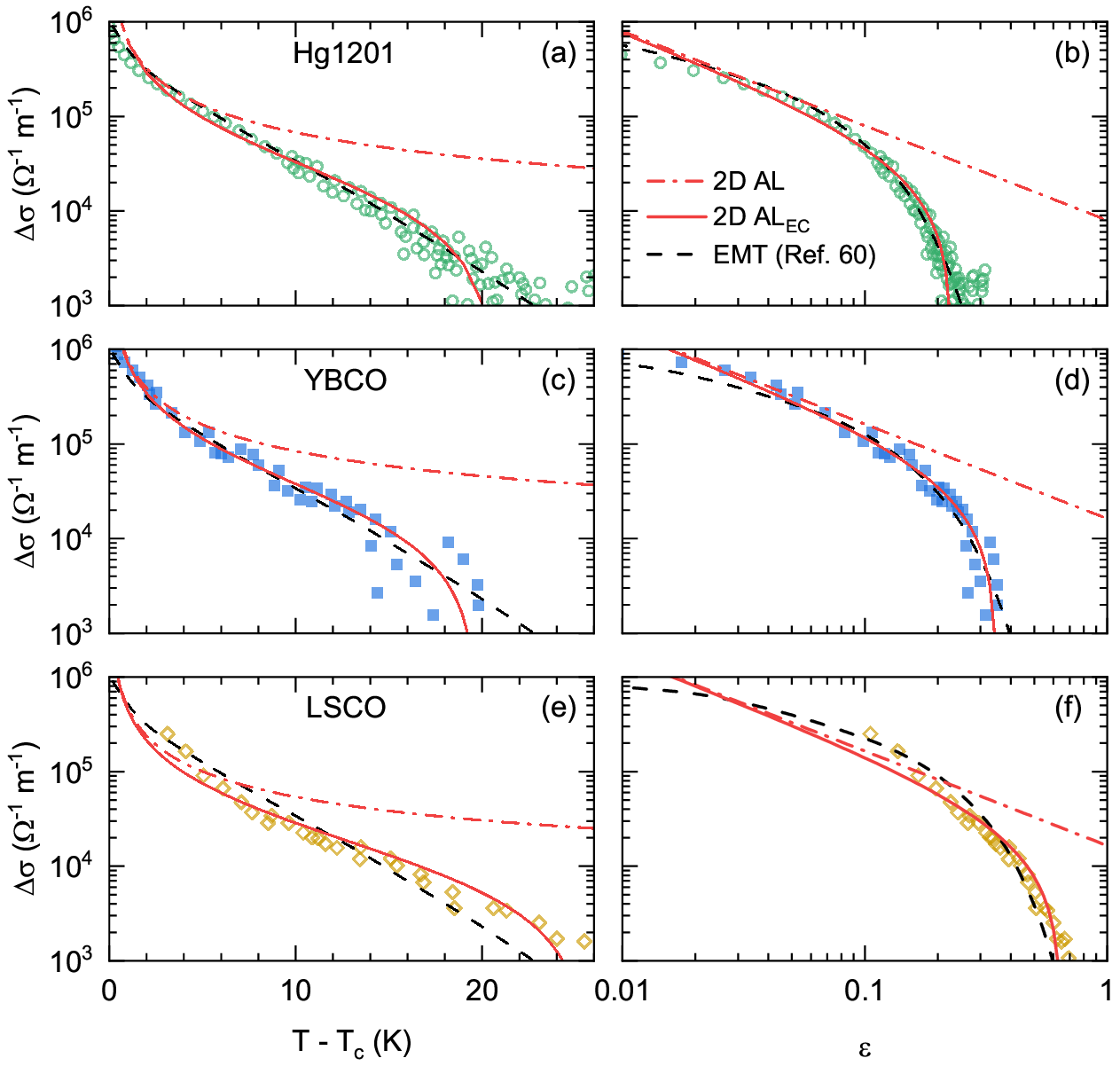}
\caption{$\Delta\sigma_{ab}(\varepsilon)$ data from Fig.~\ref{fig:Fig4}, shown separately for each compound as a function of both $T-T_c$ and $\varepsilon$. Each data set is shown next to curves corresponding to the percolative model of Ref.~\onlinecite{Popcevic18} (denoted as EMT, dashed black line), the 2D AL result (dot dashed red line) and the 2D AL result with a total-energy cutoff (2D AL$_{\rm EC}$, solid red line). Notice the difference between the EMT approach and the LSCO data despite the amplitude adaptation introduced in Ref.~\onlinecite{Popcevic18} (see the main text for details).}
\label{fig:Fig5}
\end{figure*}

The central starting point of the analyses presented in Ref.~\onlinecite{Popcevic18} is the claim that their paraconductivity measurements \textit{demonstrate a remarkable degree of universality}. This conclusion contradicts the GGL scenarios for layered superconductors, which predict a compound-dependent paraconductivity, particularly depending on the relative Josephson coupling between superconducting layers \cite{Lawrence70,Tsuzuki71,Yamaji72,Tinkham,Larkin05,Skocpol75,Vidal-Ramallo98,Llovo-Mosqueira22,Mosqueira-Carballeira01,Carballeira-Mosqueira01,Rey-Carballeira19,Maki-Thompson89,Klemm90}. To demonstrate such universality, Fig.~2 of Ref.~\onlinecite{Popcevic18} shows $\Delta\sigma$ vs. $T-T_c$ of three underdoped cuprates: LSCO ($T_c=28$~K), YBCO $T_c=47$~K), and HgBa$_2$CuO$_{4+\delta}$ (Hg1201, $T_c=80$~K). Here we will show the shortcomings of this analysis. Notice first that the paraconductivity of LSCO was divided by a factor of two to make it scale with the ones of underdoped YBCO and Hg1201. Indeed, $\Delta\sigma$ may be appreciably affected by temperature independent factors, such as the geometrical ones associated to the finite size of the electrical contacts in small or irregular samples. These uncertainties are generally considered by introducing an adjustable, temperature-independent factor in the theory to be probed (see below). However, it is questionable to discard the AL approach (whithout any cutoff, see below) after a comparison with data that has been forced to scale with the percolative approach (as it is done in Fig.~2 of Ref.~\onlinecite{Popcevic18}, see also our Fig.~\ref{fig:Fig4}). In addition, with the amplitude factor for LSCO proposed in Ref.~\onlinecite{Popcevic18}, the seeming universality is just a consequence of the data dispersion and the small resolution of the logarithmic representation used. The actual differences concern not only the $\Delta\sigma$ amplitudes but also their $\varepsilon$-behavior, as it can be seen in Fig.~\ref{fig:Fig5}, where the data sets are shown separately for each compound as a function of both $T-T_c$ and $\varepsilon$. For instance, for $\varepsilon\stackrel{>}{_\sim}0.02$, where the data are probably not affected by $T_c$-inhomogeneities, the YBCO paraconductivity is more than two times larger than for Hg1201 (these two compounds not being affected by any amplitude correction). As it can also be seen in Figs.~\ref{fig:Fig5}~(b,d,f), the $\varepsilon$-onset of $\Delta\sigma$ varies between 0.23 and 0.65 (see below), beyond the data dispersion. 

%
%

\begin{figure}[h]
\includegraphics[width=0.4\textwidth]{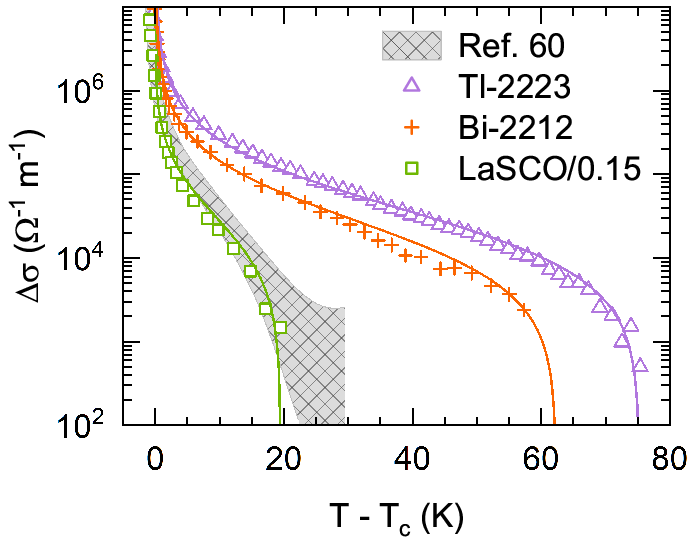}
\caption{Comparison of the $\Delta\sigma_{ab}(\varepsilon)$ data for Tl-2223, Bi-2212, and LaSCO/0.15, with the data for all the compounds studied in Ref.~\onlinecite{Popcevic18} (grey area). The lines are the GGL fits to each data set, as described in the main text. LaSCO/0.15 scales with the data from Ref.~\onlinecite{Popcevic18}, but the also optimally-doped Bi-2212 and Tl-2223 data do not. This rules out the universal $\Delta\sigma$ behavior proposed in the percolative scenario of Ref.~\onlinecite{Popcevic18}.}
\label{fig:Fig6}
\end{figure}

The non-universality of the scaling proposed in Ref.~\onlinecite{Popcevic18} is further evidenced when other samples are added to the same $\Delta\sigma(T-T_c )$ representation. Figure~\ref{fig:Fig6} shows our paraconductivity data for optimally-doped Tl2223, Bi2212 and LSCO, superimposed on the data sets from Fig.~2 of Ref.~\onlinecite{Popcevic18} (shown as a gray area for clarity). The differences between our data sets are well beyond the experimental uncertainties, even if the paraconductivity amplitude of our data were adapted as proposed in Ref.~\onlinecite{Popcevic18}. Nevertheless, our LaSCO/0.15 data are compatible with the LSCO data from Ref.~\onlinecite{Popcevic18}.

\begin{figure}[h]
\includegraphics[width=0.4\textwidth]{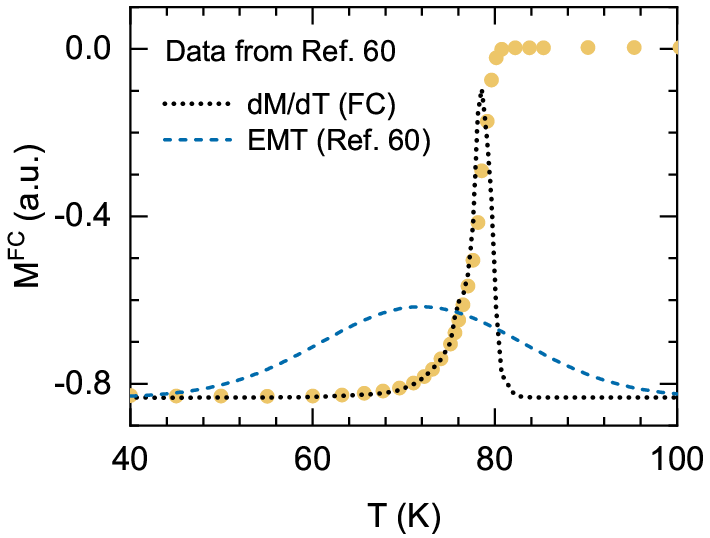}
\caption{Field-cooled magnetization of the Hg1201 sample, captured from Fig.~1\textbf{c} of Ref.~\onlinecite{Popcevic18}. As explained in the main text, $dM/dT$ (black dotted line, a.~u.) approximates the actual $T_c$ distribution for this sample. This distribution strongly contrasts with the one resulting from the percolative model proposed in Ref.~\onlinecite{Popcevic18} (blue dashed line, a.~u.). }
\label{fig:Fig7}
\end{figure}

A quantitative analysis shows further difficulties with the EMT model proposed in Ref.~\onlinecite{Popcevic18}. In their approach, the only free parameter was the full width at half maximum $\Delta T_c$ of the Gaussian $T_c$ distribution (see Ref.~\onlinecite{Popcevic18} for the details). The best fit to their data set is shown in Fig.~2 of Ref.~\onlinecite{Popcevic18} and is reproduced in Fig.~\ref{fig:Fig4} here. It leads to $\Delta T_c= 26\pm1$~K, which is a significant fraction of the $T_c$ value of the samples studied (28~K for LSCO, 47~K for YBCO, and 80~K for Hg1201). Such a large $\Delta T_c$ is inconsistent with the low-field (15~Oe) field-cooled magnetization $M^{FC}$ from Ref.~\onlinecite{Popcevic18}, which is temperature independent up to $\sim10$~K below $T_c$. A wide $T_c$ distribution would lead to a temperature dependent $M^{FC}$ well below $T_c$ (some relevant examples from the early times of high-$T_c$ cuprates may be seen in Refs.~\onlinecite{Chi_Obradors, Chi_Schilling}) In fact, the actual $T_c$ distribution may be estimated from the temperature derivative of $M^{FC}$ (see e.g., Ref.~\onlinecite{Vidal-Veira01}). The $T_c$ distribution for Hg1201 obtained with this procedure is shown in Fig.~\ref{fig:Fig7}, the resulting $\Delta T_c$ being $\sim3$~K, one order of magnitude smaller than the one resulting from the percolative approach fit of Ref.~\onlinecite{Popcevic18}, also represented in Fig.~\ref{fig:Fig7}. 

Another point emphasized in Ref.~\onlinecite{Popcevic18}, is that \textit{the exponential temperature dependence} (shown by their paraconductivity data) \textit{is incompatible with standard models of superconducting fluctuations such as Ginzburg-Landau theory}. To arrive to this conclusion, a comparison of data for different compounds with the 2D AL expression without any cutoff \cite{Aslamazov68} was made in Fig.~2 of Ref.~\onlinecite{Popcevic18}. However, as we have already stressed in section III.C, the original AL paraconductivity overestimates the short wavelength fluctuating modes, and a momentum cutoff was early introduced to mitigate such a disagreement at high reduced temperatures \cite{Tinkham,Larkin05,Skocpol75,Johnson78}. In addition, a single curve, calculated by using only the Hg1201 effective interlayer distance, was used in Fig.~2 of Ref.~\onlinecite{Popcevic18}, even though the 2D AL paraconductivity is dependent on the effective interlayer distance and on $T_c$.

%
%

%
%

For the aforementioned reasons, we will analyze Ref.~\onlinecite{Popcevic18} data in terms of the AL paraconductivity for 2D layered superconductors up to the high-$\varepsilon$ region by using the GGL approach with a total energy cutoff \cite{Tinkham,Larkin05,Skocpol75,Vidal-Ramallo98,Curras-Ferro03,Johnson78,Freitas87,Lee-Klemm89,Gauzzi95}. The possible amplitude indeterminations affecting $\Delta\sigma_{ab}$ from Ref.~\onlinecite{Popcevic18} were considered by including a temperature-independent multiplicative parameter $C_g$ in Eq.~(\ref{eq:delta_sigma_2D}). Only data above $T_c$ and up to the $\varepsilon$-value at which $\Delta\sigma_{ab}$ falls below the noise level were used, resulting in a $\Delta\sigma_{ab}$ window in Fig.~\ref{fig:Fig5} between $10^3\;\Omega^{-1}$m$^{-1}$ and $10^6\;\Omega^{-1}$m$^{-1}$ (for $\Delta\sigma_{ab}$ below this value the signal-to-noise ratio becomes too low). The cutoff parameters were estimated from Figs.~\ref{fig:Fig5}~(b,d,f) as the reduced temperatures at which $\Delta\sigma_{ab}\approx10^3\;\Omega^{-1}$m$^{-1}$. 

The 2D AL curve for each compound is shown in Fig.~\ref{fig:Fig5} as a dot-dashed line, and the curve from Eq.~(\ref{eq:delta_sigma_2D}) with the $C_g$ values indicated in Table~\ref{Table4} as solid red lines (the curves obtained for Hg1201 are also shown in Fig.~\ref{fig:Fig4}). As it can be seen, the agreement with the experimental data is remarkably good even in the high-$\varepsilon$ region. Nevertheless, the observed differences between the $\varepsilon^c$ values for the different compounds may be associated with differences in the relationship between the Pippard and the GGL coherence length amplitudes \cite{Vidal-Carballeira02,Vidal-Ramallo03}, or to uncertainties associated to the background subtraction procedure, as the background determination is particularly difficult in non-optimally-doped compounds \cite{Vidal-Veira01,Curras-Ferro03}.

Finally, it is worth noting two points. First, the adequacy of the superconducting fluctuations scenario in cuprates was also earlier supported by the analyses of the behavior above but near $T_c$ of other observables, mainly the in-plane magnetization (see, e.g., Refs.~\cite{Vidal-Ramallo98,Rey-Carballeira19,Mosqueira-Dancausa11,Borna22,Maignan94}). Second, the $\Delta\sigma_{ab}$ measurements in underdoped LSCO under magnetic fields up to 50~T presented in Ref.~\cite{Leridon07} are also consistent with the 2D-AL approach near $T_c$, and show a reduction at high-$\varepsilon$ that is related to the total-energy cutoff (see also Ref.~\cite{Caprara05}). The $\Delta\sigma_{ab}$ data in slightly underdoped YBCO also show a good agreement with the 3D AL approach near $T_c$, and the seemingly exponential behavior at higher temperatures seems to be a crossover to the cutoff-dominated region\cite{Leridon01}.

%
%
\begin{table}[h]
\begin{ruledtabular}
\begin{tabular}{ccccccccc}
Sample & $d$  & $N$ & $T_c$ & $\Delta T_c$ & $d_{\rm eff}$ & $C_g$ & $\varepsilon^c$ & $\xi_{ab}(0)$\\
 & (nm) &   & (K) & (K) & (nm) &  &   & (nm)\\
\hline
Hg1201 & 0.95 & 1 & 80 & 0.8$\pm$0.2 & 0.95 & 0.5 & 0.23  & 3.0\\
YBCO & 1.17 & 2 & 47 & 2.2$\pm$0.2 & 0.585 & 0.62 & 0.35  & 2.9\\
LSCO & 0.66 & 1 & 28 & 1.8$\pm$0.2 & 0.66 & 0.71 & 0.65 & 3.5\\
\end{tabular}
\end{ruledtabular}
\caption{Parameters corresponding to the analysis of the paraconductivity data of Ref.~\onlinecite{Popcevic18} in terms of the 2D AL approach with a total-energy cutoff, Eq.~(\ref{eq:delta_sigma_2D}). $\Delta T_c$ was estimated as the FWHM of $d\rho_{ab}/dT$. A multiplicative constant parameter $C_g$ was also included to account for the uncertainties in the $\Delta\sigma$ amplitude (due, in particular, to possible uncertainties in the samples geometry). Without the amplitude adaptation introduced in Ref.~\onlinecite{Popcevic18}, for the LSCO compound $C_g$ would be $\sim 1.4$. The resulting curves can be seen in Fig.~\ref{fig:Fig4} for Hg1201 and Fig.~\ref{fig:Fig5} for all compounds (red solid curves). For completeness, values of the in-plane coherence length amplitude taken from Refs.~\onlinecite{Hofer98,Gray92,Mosqueira-Cabo09} are also included.}
\label{Table4}
\end{table}

\section{Conclusions}

The $\rho_{ab}(T)$ roundings measured just above $T_c$ in three optimally-doped compounds La$_{1.85}$Sr$_{0.15}$CuO$_4$, Bi$_2$Sr$_2$CaCu$_2$O$_{8+\delta}$, and Tl$_2$Ba$_2$Ca$_2$Cu$_3$O$_{10}$ \cite{Curras-Ferro03,Campa92,Maignan94,Vina02}, have been used for a throughout confrontation between percolation processes and the unavoidable presence of fluctuating superconducting pairs. The excellent chemical and structural quality of these optimally-doped samples minimizes the effect of extrinsic $T_c$-inhomogeneities, a crucial aspect when analyzing the possible presence of intrinsic percolative processes. In addition, these compounds have different number of superconducting $ab$-layers in their periodicity length, and different Josephson coupling between layers, which is particularly relevant to probe the presence of fluctuating superconducting pairs. In our analyses, we have excluded the so-called full-critical region close to $T_c$ (a reduced temperature region also affected, mainly in the case of the LaSCO/0.15 sample, by $T_c$-inhomogeneities). However, we have covered the high-$\varepsilon$ region, up to the resistivity rounding onset.

A remarkable result of our analysis is the agreement between the $\varepsilon_{\rm onset}$ values well within the experimental uncertainties. These $\varepsilon_{\rm onset}$ are also consistent with the value (0.55) predicted by the Gaussian-Ginzburg-Landau approach under a total-energy cutoff for clean superconductors \cite{Vidal-Carballeira02,Vidal-Ramallo03}. The latter was introduced heuristically \cite{Vidal-Carballeira02,Vidal-Ramallo03}, as it also was the classical momentum cutoff \cite{Tinkham,Skocpol75,Johnson78}. A similar $\varepsilon_{\rm onset}$ was also found when analyzing the $\rho_{ab}(T)$ rounding and the precursor diamagnetism above $T_c$ in optimally-doped YBCO \cite{Rey-Carballeira19}. These results confirm earlier results in different optimally-doped cuprates \cite{Vidal-Carballeira02,Vidal-Ramallo03,Curras-Ferro03,Vina02} at a quantitative level, evidencing that the unavoidable fluctuating superconducting pairs must be the origin of the precursor conductivity in optimally-doped cuprates.

Additionally, by using the simplest form of the effective-medium theory we conclude that percolative processes alone cannot account for the measured $\Delta\sigma_{ab}(\varepsilon)$, generalizing previous findings in optimally-doped YBCO \cite{Maza91} (see also the short conference proceedings recently published in SN Applied Science, Ref.~\onlinecite{Llovo-Mosqueira22}). In contrast, these measurements have been quantitatively explained using the GGL approach for layered superconductors, extended to high-$\varepsilon$ by including the total-energy cutoff. In the case of Bi$_2$Sr$_2$CaCu$_2$O$_{8+\delta}$ and Tl$_2$Ba$_2$Ca$_2$Cu$_3$O$_{10}$, for which the superconducting fluctuations in the $0.01 < \varepsilon < 0.1$ range are in the 2D limit, the analysis has been done without adjustable parameters. In the case of optimally-doped cuprates, our results provide an unambiguous conclusion to the question addressed by Bednorz and M\"uller in their seminal work \cite{Bednorz86}: The enhanced conductivity observed above but near $T_c$ can be quantitatively accounted for, up to the rounding onset temperature, by the presence of unavoidable superconducting pairs created by the thermal agitation energy. This conclusion is also supported by the analyses of the behavior above but near $T_c$ of other observables, mainly the in-plane magnetization \cite{Vidal-Ramallo98,Rey-Carballeira19,Mosqueira-Cabo09,Mosqueira-Dancausa11,Borna22,Maignan94}.

Finally, we have shown that data from a recent work seemingly supporting a percolative scenario in three underdoped cuprates can in fact be explained on the grounds of conventional thermal fluctuations. This result is not trivial, because these materials are expected to be affected to some extent by $T_c$ inhomogeneities: their $T_c$ is strongly dependent on the doping level, contrary to optimally doped cuprates. On the other hand, the doping level presents a spatial variation due to the random distribution of dopants and to the small $\xi_{ab}$ amplitude (of the order of nm). Our analysis suggests that, in these underdoped samples, $T_c$ inhomogeneities play a non-negligible role only in a restricted temperature range (just a few K around $T_c$) where the AL approach deviates appreciably from the data. These results reinforce the adequacy of fluctuation models for the conductivity enhancement above but not too close to $T_c$, and directly address the proposals of other authors questioning the GGL scenarios for the paraconductivity in cuprates, particularly in the high-$\varepsilon$ region.

\begin{acknowledgments}
We thank C. Carballeira, J. Maza, and M. V. Ramallo for earlier discussions on the theoretical approaches for the paraconductivity. This work was supported by the Agencia Estatal de Investigaci\'on (AEI) through project PID2019-104296GB-I00. I.~F. Llovo acknowledges financial support from Xunta de Galicia through grant ED481A-2020/149. 
\end{acknowledgments}
\textbf{Conflict of interest statement:} The authors declare that there is no conflict of interest to disclose.

\textbf{Data access statement:} Data will be made available upon reasonable request.

\end{document}